\documentclass[twocolumn,showpacs,showkey,nofootinbib]{revtex4}
\usepackage{graphicx}
\usepackage{soul}
\usepackage{color}

\usepackage{amsmath}
\usepackage{amssymb}
\usepackage{pgf}
\usepackage{subfigure}
\definecolor{OliveGreen}{rgb}{0,0.6,0}

\newcommand{\be}{\begin{eqnarray}}
\newcommand{\bel}{\begin{equation}\label}
\newcommand{\ee}{\end{equation}}
\def\dg{\dagger}
\newcommand{\barl}{\begin{eqnarray}\label}
\newcommand{\ear}{\end{eqnarray}}
\newcommand{\non}{\nonumber}
\begin{document}
%%%%%%%%%%%%%%%%%%%%%%%%%%%%%%%%%%%%%%%%%%%%%%%%%%%%%%%%%%%%%%%%%%%%%%%%%%%%%%%%
\title{Electromagnetic pulse transparency in coupled cavity arrays through dispersion management}
%%%%%%%%%%%%%%%%%%%%%%%%%%%%%%%%%%%%%%%%%%%%%%%%%%%%%%%%%%%%%%%%%%%%%%%%%%%%%%%%
\author{Z. Ivi\'c$^{1,2}$, Z. Pr\v zulj$^1$, D. Chevizovich$^1$, 
G. P. Tsironis$^{2,3}$}
%%%%%%%%%%%%%%%%%%%%%%%%%%%%%%%%%%%%%%%%%%%%%%%%%%%%%%%%%%%%%%%%%%%%%%%%%%%%%%%%
\affiliation{$^{1}$ Laboratory for Theoretical and Condensed Matter Physics, "VIN\v CA" Institute of Nuclear Sciences, National Institute of the Republic of Serbia, University of Belgrade P.O.Box 522, 11001 Belgrade, Serbia,\\
$^{2}$Institute of Theoretical and Computational Physics, Department of Physics, 
The University of Crete, P.O. Box 2208, Heraklion, 71003, Greece \\
$^{3}$Institute of Electronic Structure and Laser,
Foundation for Research and Technology--Hellas, P.O. Box 1527, 71110 
Heraklion, Greece \\
}
%%%%%%%%%%%%%%%%%%%%%%%%%%%%%%%%%%%%%%%%%%%%%%%%%%%%%%%%%%%%%%%%%%%%%%%%%%%%%%%%
\date{\today}
%%%%%%%%%%%%%%%%%%%%%%%%%%%%%%%%%%%%%%%%%%%%%%%%%%%%%%%%%%%%%%%%%%%%%%%%%%%%%%%%

%%%%%%%%%%%%%%%%%%%%%%%%%%%%%%%%%%%%Abstract-Here%%%%%%%%%%%%%%%%%%%%%%%%%%%%%%%
\begin{abstract}
We theoretically demonstrated the possible emergence of slow-light self-induced transparency solitons in the infinite one-dimensional coupled cavity array, with each cavity containing a single qubit. We have predicted a substantial dependence of pulse transparency on its dimensionless width -- $\tau_0$. In particular, short pulses whose widths range from $\tau_0\ll 1$ to $\tau_0\lesssim 1$ exhibit simple, almost linear dispersion law with a finite frequency gap of the order of the cavity array photonic band gap. That is, the medium is opaque for very short pulses with carrier wave frequency below the photonic gap. When the pulse width exceeds the critical one, a twin transparency window separated by a finite band gap appears in the soliton pulse dispersion law.
Observation of predicted effects within the proposed setup would be of interest for understanding the properties of self--induced transparency effect in general and future applications in the design of quantum technological devices.
\end{abstract}
%%%%%%%%%%%%%%%%%%%%%%%%%%%%%%%%%%%%%%%%%%%%%%%%%%%%%%%%%%%%%%%%%%%%%%%%%%%%%%%%
\maketitle
%%%%%%%%%%%%%%%%%%%%%%%%%%%%%%%%%%%%%%%%%%%%%%%%%%%%%%%%%%%%%%%%%%%%%%%%%%%%%%%%

%%%%%%%%%%%%%%%%%%%%%%%%%%%%%%%%%%%%%%%%%%%%%%%%%%%%%%%%%%%%%%%%%%%%%%%%%%%%%%%%
\section{Introduction}
The achievement of control over the propagation of
electromagnetic radiation in engineered media comprised
of a large number of artificial two-level atoms (qubits)
is a prerequisite for the practical realization of operable
devices for quantum information processing and communication.
Specifically, light slowing down, even stopping, is of vital interest for the realization of devices for long -- lived quantum memories \cite{cqp1,cqp2,cpq3,cpq4,appl,QIT1,QIT4,eit1,eit2,eit3,storage1,storage2,qusol1,qusol2,qusol3,bull,aga,slowqmm}. Natural candidates
for quantum information storage and retrieval are
slow-light quantum optical solitons representing multiphoton
bound states \cite{eit1,eit2,eit3,storage1,storage2,qusol1,qusol2,qusol3}.They may emerge either due to strong photon-photon interaction \cite{qusol1} or polaritonic effect. In the latter case, slow light can be generated employing: electromagnetically induced transparency (EIT) and self--induced transparency, for example (SIT) \cite{eit1,eit2,eit3,qusol2,qusol3}. Great practical success in slow light production has been achieved by using EIT-based techniques. Nevertheless, in most cases EIT and optical pulse propagation have been performed using atomic gases in bulk samples in which light-matter interaction is limited while the EIT effect is weak. All this implies the necessity for the quest of novel methods, non--EIT based ones, of slow light production \cite{aga,slowqmm}.

The construction of the platform for the generation of stable and controllable QOS represents a necessary step toward their practical usage. Recently  \cite{scir,flux,disp}, we have shown that QOSs, exhibiting typical features of SIT pulses in atomic gases \cite{sit1,sit2,sit3,sit4,sit5} may emerge in quantum metamaterials (QMM): one--dimensional (1d) arrays of superconducting qubits embedded in superconducting transmission lines \cite{qmm1,qmm2,qmm3,qmm4,qmm5,qmm6}. The adjustability of qubit parameters provides control over the strength of mater-light interaction and soliton propagation. We recall that SIT is a coherent, lossless propagation of short, powerful, electromagnetic pulses through a medium built of \textit{inhomogeneously broadened} (IHB) two-level atoms, provided that the pulse areas are equal to integer multiples of $\pi$: $n\pi, \; n=1,2,...$. Note that IHB is a prerequisite for the emergence of SIT \cite{sit1,sit2,sit3,sit4,sit5}. That means that by employing the SIT effect, it could be possible to exploit relaxation phenomena, IHB in particular, to support achievement and maintain quantum coherence, which could be of interest for further development of quantum technologies.

In this article, we study a possible emergence of SIT-like solitons in QMM comprised of an infinite 1d chain of optical cavities uniformly "doped" with a large number of qubits and embedded in 2d photonic crystal \cite{yariv,cca,ccapolarit}. Due to the tunability of qubit parameters and possible engineering of the periodic structure of photonic crystals, such media could be a flexible platform to support solitons usable for information storage and manipulation.

\section{Setup and mathematical model}
\begin{figure}[h]
\includegraphics[height=4.5cm]{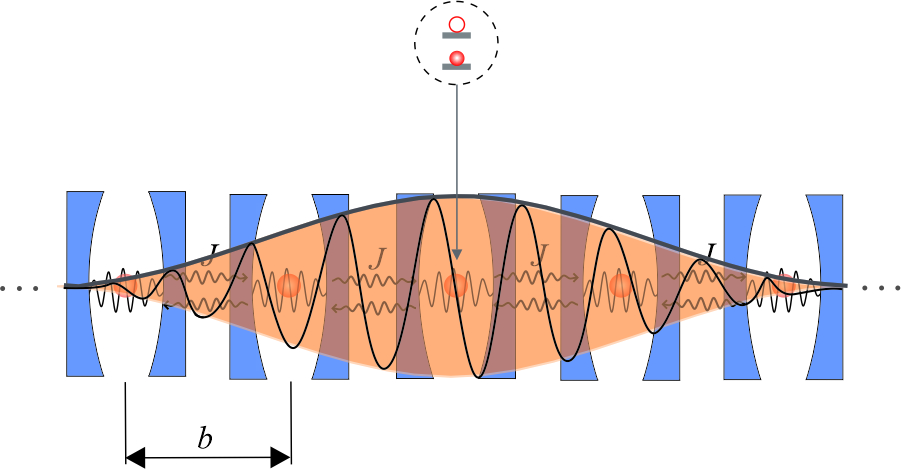}\caption{ Schematic representation of an infinite 1d array of coupled single--mode cavities, each containing a single qubit. The shaded region depicts propagating electromagnetic pulse. A wavy, fast-oscillating line indicates the carrier wave. }\label{fig04}
\end{figure}

The system under consideration is illustrated in ( Fig. \ref{fig04}). It is a 1d waveguide comprised of coupled, identical, equally spaced at distance - $b$, single-mode optical cavities, each filled with single qubit \cite{cca,ccapolarit}.
The type of qubit, irrelevant in the present context, should be specified in the practical realization of a particular quantum technological device. We only assume that all qubits are ideally symmetric (unbiased) two-level systems. In realistic conditions, bias is always present but may be eliminated by an appropriate choice of external parameters: the magnetic field for flux qubits or gate voltage for charge qubits. In each cavity, qubits strongly interact with a quantized electromagnetic field with frequency $\omega_0$. Light propagation along such a structure is achieved through photon tunneling between adjacent units characterized by the inter-cavity tunneling energy $J$. The Hamiltonian of the system is Bose tight--binding Hamiltonian:
\begin{eqnarray}\label{h1}
\non H=&&\sum_n\left[\frac{\hbar\Delta}{2}\sigma^z_n +\hbar\omega_0 a^{\dg}_n a_n+\hbar \frac{g}{2}\sigma^x_n(a^{\dg}_n+a_n)\right]-
\\&&J\sum_na^{\dg}_n (a_{n+1}+a_{n-1}).
\end{eqnarray}

The single-photon coupling constant with qubit in the same cavity is $g$. $a^{\dg}_n$ ($a_n$) is photon creation (annihilation) operator in $n$--th cavity; finally, qubit operators, specified through the Pauli matrices, in the representation of basis $|g\rangle; |e\rangle$, read: $\sigma^{x}_n=|g\rangle_n\langle e| + |e\rangle_n\langle g|$, $\sigma^{y}_n=i(|e\rangle_n\langle g| -|g\rangle_n\langle e|$) and $\sigma^{z}_n=|e\rangle_n\langle e| - |g\rangle_n\langle g|$. At each site "particle" may be either in the ground or excited state, implying the following constraint: $|e\rangle_n\langle e| + |g\rangle_n\langle g|=1$, which is consistent with the $\sum_{l=x,y,z}(\sigma^l_n)^2\equiv 1$. 

To simplify practical work, avoid technical problems, and ensure accurate calculations, we first rewrite (\ref{h1}) in reciprocal space employing Fourier transform of photon operators $a_n=\frac{1}{\sqrt{N}}\sum_q e^{iqnb} a_q $. This yields:
\begin{eqnarray}\label{hft}
\non && H=\sum_q\hbar\omega_q a^{\dg}_q a_q+\hbar\sum_{n,q}\frac{g^q_n}{2}\sigma^{x}_n(a_q+a^{\dg}_{-q})+\frac{\hbar\Delta}{2}\sum_n\sigma^z_n\\
&&\omega_q=\omega_0-\frac{2J}{\hbar}\cos qb,\;\; g^q_n=\frac{g e^{iqnb}}{\sqrt{N}}.
\end{eqnarray}

The quantum state of any system evolves in time according to Schr\"odinger equation $i\hbar\frac{\partial}{\partial t}|\Psi\rangle = H|\Psi\rangle$. 
To derive desired evolution equations, we employ the time-dependent variational principle. Within the semiclassical approximation \cite{scir,disp, flux}, we chose the vector of a state as a direct product of vectors of states of atom and photon subsystems:
\begin{eqnarray}\label{vs}
|\Psi(t)\rangle=\sum_{n;p=g,e}\varPsi^p_{n}(t)|p\rangle_n\otimes\prod_q|\alpha_q(t)\rangle.
\end{eqnarray} 
Here $|\alpha_{q}(t)\rangle$ is the single mode photon coherent state defined as eigen--state of photon annihilation operator: $a_q|\alpha_{q}(t)\rangle=\alpha_q(t)|\alpha_q(t)\rangle$.
For convenience, to simplify some mathematical steps, we did not impose normalization constraint on (\ref{vs}). Nevertheless, due to the aforementioned algebraic properties of Pauli matrices, $\sum_{l=x,y,z}(\sigma^l_n)^2\equiv|e\rangle_n\langle e| + |g\rangle_n\langle g|=1$, for each site, say some arbitrary $n_0$, (\ref{vs}) satisfies:
\begin{equation}\label{norm}
\langle \Psi(t)|(|e\rangle_{n_0}\langle e|+|g\rangle_{n_0}\langle g|)|\Psi(t)\rangle=|\varPsi^e_{n_0}|^2+|\varPsi^g_{n_0}|^2\equiv 1.
\end{equation}

Evolution equations for $\varPsi_{p}(t)$ and $\alpha_q(t)$ follow requiring stationarity of Lagrange action
$L=\int_{t_1}^{t_2} dt \mathcal{L}$, where $\mathcal{L}=\frac{i\hbar}{2}\left(\langle \Psi|\dot\Psi\rangle-c.c\right)-\langle\Psi|H|\Psi\rangle$ denotes Lagrangian density. Thus, demanding $\delta L=0$ we easily obtain Hamilton--like equations:
\begin{eqnarray}\label{hamj}
i\hbar \dot\varPsi^p_n=\frac{\partial \mathcal{H}}{\partial \varPsi^*_{p}}, \;\;\; i\hbar \dot\alpha_q=\frac{\partial \mathcal{H}}{\partial \alpha^*_{q}},
\end{eqnarray}
where, $\mathcal{H}=\langle \Psi(t)|H|\Psi(t)\rangle$ plays the role of the classical Hamiltonian function. 

Employing (\ref{hamj}), we obtain the following system of evolution equations:
\begin{eqnarray}\label{se}
\non i\dot\varPsi^e_n=\frac{\Delta}{2}\varPsi^e_n+\sum_{q}\frac{g^q_n}{2}(\alpha_q+\alpha^{*}_{-q})\varPsi^g_{n}\\
i\dot\varPsi^g_n=-\frac{\Delta}{2}\varPsi^g_n+\sum_{q}\frac{g^q_n}{2}(\alpha_q+\alpha^{*}_{-q})\varPsi^e_{n},\\ \non
i\dot\alpha_{q}=\omega_q\alpha_{q}+\sum_n\frac{g^{-q}_n}{2}(\varPsi^e_n\varPsi^{g*}_{n}+c.c),
\end{eqnarray}
which, together with the corresponding ones for their complex conjugate, fully describe the system dynamics. Here we have neglected the effects of dissipation 
(homogeneous broadening), which are irrelevant for the (ultra)short pulses whose duration time is short compared with both longitudinal and transverse relaxation times \cite{sit1,sit2,sit3,sit4,sit5}. 

Similar equations have been derived recently in the studies of the polariton--soliton \cite{sol4,sol5} and photonic rogue waves \cite{sol3}  in  CCAs filled with the ensembles of qubits. It was found that the competition between the photon tunneling and nonlinearities arising from the coupling of cavities with ensembles of mutually interacting qubits may lead to soliton formation. \cite{sol4,sol5}. Finally, an additional nonlinearity coming from photon--photon coupling gives rise to creation of rogue waves \cite{sol3}. 

In the present work, we examine whether equations (\ref{se}) possess soliton solutions exhibiting lossless propagation consistent with the \textit{area theorem} \cite{sit1,sit4,sit5}. For that purpose, in the appendix (\ref{MB}), we have proved their equivalence with the system of Maxwell -- Bloch equations. Then, using a standard procedure, continuum, and slowly varying envelope and phase approximations, we, in the appendix (\ref{RMBE}), have derived a system of reduced Maxwell - Bloch equations (RMBE) (\ref{rmbe1}) and (\ref{rmbe2}). Their analytic traveling wave solutions are well known (see appendix (\ref{tvsol})), implying the possible emergence of SIT within the present model. For a demonstration of the SIT pulse lossless propagation, we have employed equivalence between the RMBEs and sine--Gordon (SG) (see appendix \ref{tga}) equation for Bloch angle ($\theta(x,t)$) to derive the energy balance equation (see appendix \ref{enbal} equation \ref{lossless1}): 
\begin{eqnarray}\label{lossless}
\frac{\partial W(x)}{\partial x}=-\mathcal{S}^0\frac{F(D)g\Omega }{kc^2_0}\bigg(1-\cos\Theta(x)\bigg),
\end{eqnarray} 
Here, $\Theta(x)$ is the so-called pulse area defined through the Bloch angle and the components of a Bloch vector by equations (\ref{angle11}) and (\ref{angle22}). While $W(x)\sim \int_{-\infty}^{\infty} dt u^2(x,t)$ denotes pulse energy density, which, averaged over the overall phase, reads $W(x)\sim \frac{1}{2}\int_{-\infty}^{\infty}dt f^2(x,t)$.
The meaning of the parameters are: $\mathcal{S}_0=\langle\Psi(-\infty)|S^z_n|\Psi(-\infty)\rangle\equiv |\varPsi^e_{n}|^2-|\varPsi^g_{n}|^2$ the initial value for population inversion. It takes values $\mp 1$ depending on whether all "atoms" are in the ground (absorbing media) or excited (amplifying media) state. $F(D)=1/(1+(D\tau_p)^2)$ is a spectral function, $\omega$ and $k$ are the frequency and wavevector of the carrier wave, while $\Omega=\omega_0-2J/\hbar$, $c_0=\sqrt{2J\Omega b^2/\hbar}$ and $D=\Delta -\omega$ are, respectively, the gap in the photon spectrum, speed of light in empty coupled cavity array and detuning between the qubit and carrier wave frequencies. 

Apparently, for specific values of pulse area $\theta = n2\pi$, there is no pulse loss or amplification. Nevertheless, according to the last result, a comprehensive study of SIT pulse propagation requires the explicit knowledge of SIT pulse dispersion law - $k(\omega; \tau_0)$.
\section{dispersive properties}
We now examine carrier wavevector ($k$) dependence on the values of system parameters which, due to (\ref{lossless}), determines SIT pulse transparency. 

For reference, we first consider the \textit{linear dispersion} corresponding to plane wave solutions which we found setting to zero all derivatives in (\ref{rmbe1}) and (\ref{rmbe2}). Straightforward calculations yield: 
\begin{eqnarray}
\non &&\mathcal{S}^x(x,t)= -\frac{\left(\Omega^2+c^2_0k^2-\omega^2\right)}{\Omega g}f(x,t), \\ && \mathcal{S}^z=-\frac{2D}{g^2\Omega}\left(\Omega^2+c^2_0k^2-\omega^2\right),\; \mathcal{S}^y=0.
\end{eqnarray}

Combining the last result with the conservation law $\sum_{l=x,y,z}(\mathcal{S}^l_n)^2\equiv 1$, in the limit $f\rightarrow 0$ and for absorbing medium, i.e. taking that initially all qubits are in ground state ($\mathcal{S}^z(t=-\infty)\approx -1$), we obtain a typical polariton--like dispersion law: 
\begin{eqnarray}\label{line}
K^2=X^2-\alpha^2+\frac{h^2}{2}\frac{\alpha}{1-X}.
\end{eqnarray}
For convenience, here we have introduced dimensionless parameters: carrier wave quasi--momentum $K= \frac{c_0 k}{\Delta}$, frequency $X=\frac{\omega}{\Delta}$, new coupling constant $h=\frac{g}{\Delta}$ and 
normalized frequency gap of the empty waveguide $\alpha=\frac{\Omega}{\Delta}$.

\begin{figure}[h]
\includegraphics[height=6.75cm]{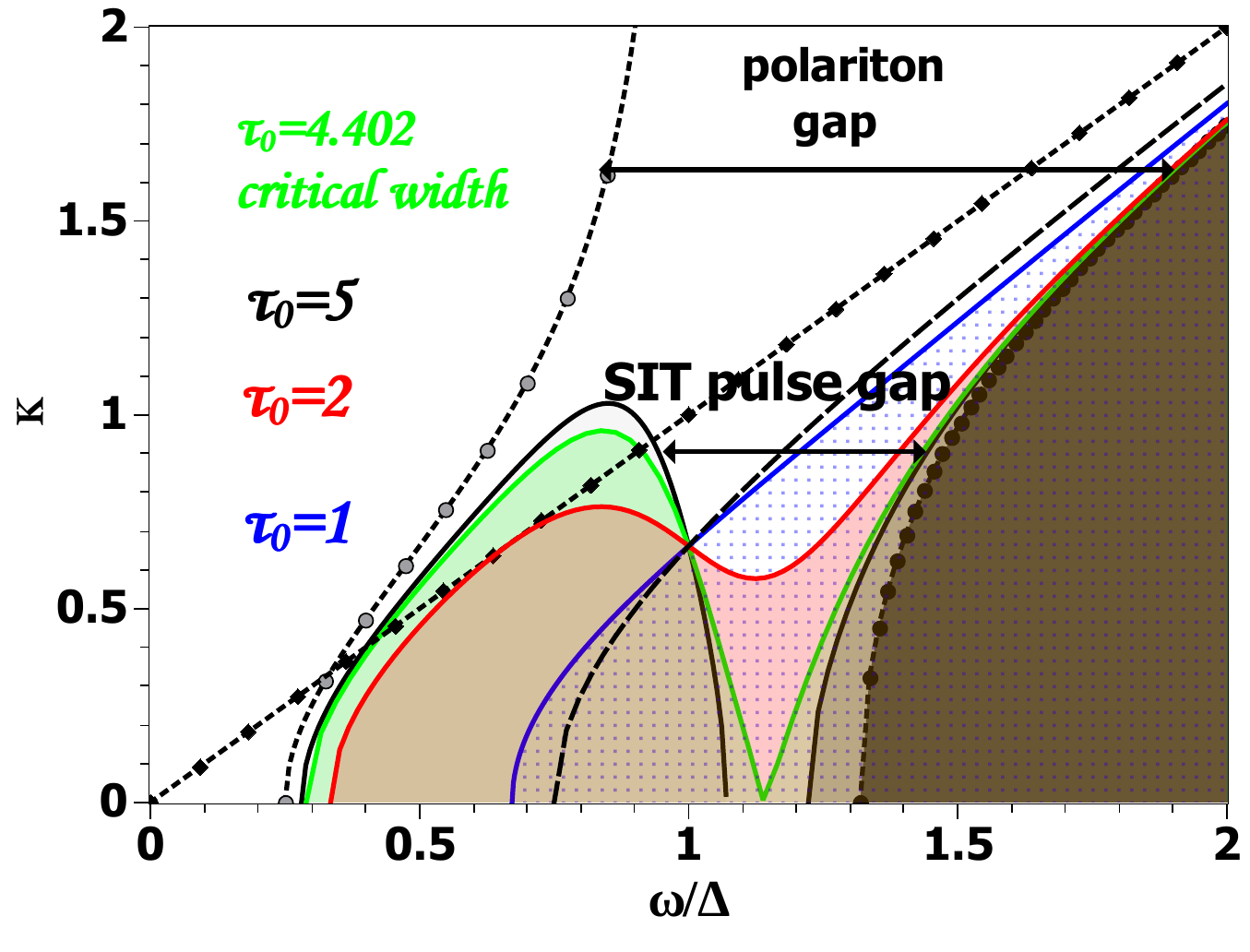}
\caption{SIT pulse dispersion spectrum for $\tau_0=1,\; 2,\;4.402\; \mathrm{and}\;5$. Shaded areas denote \textit{transparency windows}. The dashed line depicts photon dispersion in the empty CCA. For the comparison, we have presented the simple linear dispersion encountered in the analysis of SIT in atomic gases \cite{sit1,sit2,sit3,sit4} -- dotted line with diamonds, and the polariton dispersion -- two dotted lines with circles.}\label{fig1}
\end{figure}

The polariton spectrum, in parallel with the dispersion law for SIT pulse, is visualized in Fig.\ref{fig1}. It consists of two branches separated by the frequency gap in which the polariton wave--vector is imaginary. Thus, the medium is non-transparent. The width of the frequency gap increases with the magnitude of the normalized optical gap ($\alpha$). In particular, as $\alpha$ grows, the low polariton branches gradually move closer and merge for $\alpha\gg 1$. 

\textit{SIT soliton dispersion}:
To this end, we exploit equation (\ref{disp}) from which we eliminate $\gamma$ employing equation for the pulse amplitude (\ref{ampl}) and the second relation in (\ref{env}). Straightforward calculations yield the SIT soliton dispersion law in terms of the dimensionless pulse width $\tau_0=\Delta\tau_p$ and the afore specified dimensionless parameters as follows
\begin{eqnarray}\label{k}
K=\sqrt{X^2-\alpha^2-\frac{S_0\alpha h^2}{2}\frac{(1 - X)\tau^2_0}{1+(1 -X)^2\tau^2_0}}.
\end{eqnarray} 
Similarly, by virtue of the explicit form of $\gamma$, eq. (\ref{rmbe11}), we found dimensionless pulse velocity ($V=v/c_0$) as follows:
\begin{eqnarray}\label{v}
V=\frac{K}{X-\frac{S_0\alpha h^2}{4}\frac{\tau^2_0}{1+(1-X)^2\tau^2_0}}.
\end{eqnarray}
We have presented our results in terms of the pulse input parameters: pulse width ($\tau_0$) and the carrier wave frequency ($\omega$). We did not consider the impact of the variation of the coupling constant on pulse dispersion features. Namely, according to (\ref{k}), it would have practically the same consequences as the variation of $\tau_0$. Thus, we chose $h=1$ and the normalized gap as $\alpha =0.75$. The presence of a gap in the photon spectrum implies the following condition for free radiation propagation $X>\alpha$. 

In figure Fig.(\ref{fig1}), along with the polariton dispersion law (dotted black curves with circles), we have graphically presented our results for SIT dispersion law (\ref{k}) for absorbing media ($S_0=-1$). It is visualized by a set of curves $K(\omega/\Delta; \tau_0)$, all placed within the polariton gap. 
To ensure a comprehensive presentation, we took ($\tau_0$) to vary from short to very wide pulses. The spectrum of each pulse may exhibit specific features. In the short pulse limit $\tau_0 \lesssim 1$ dispersion curves, as a function of normalized pulse frequency, exhibit smooth, almost linear, increase close to that of free photons $K\sim \sqrt{X^2-\alpha^2}$ depicted through the black dotted curve. For short pulses, $K$ has real finite values for all frequencies $X\gtrsim\alpha, \; (\omega > \Omega)$ so that the medium is fully transparent. 

For wider pulses, $\tau_0\gtrsim 2$, dispersion law attains peculiar nonmonotonic behavior characterized by the emergence of two extremes, minimum and maximum, in each $K(X;\tau_0)$ curve. When pulse width overgrows some critical value ($\tau_0\sim 4.4$ for the present choice of $\alpha$, a green line at Fig. (\ref{fig1})), each curve splits into the two branches and SIT pulse spectrum consists of three bands. Two of them, corresponding to finite real $K$, constitute a twin transparency window (TW) -- dark shaded areas at Fig. (\ref{fig1}). Transparency windows are separated by a forbidden band whose wideness increases with the pulse width $\tau_0$. The low-frequency window and forbidden band both lie entirely within the polariton gap. Low-frequency TW represents a closed area in the $K-X$ plane whose lower boundary merges with the corresponding one of the polaritonic band, while its upper boundary gradually tends to the line $X\approx 1$ with the rise of pulse width $\tau_0$. At the same time, $K$ cannot exceed some maximal value determined by $\tau_0$. 
On the contrary, high-frequency TW is much like the high-frequency polariton branch. Their boundaries are very close and finally merge for wide pulses ($\tau_0\sim 10$). For the frequencies falling out of TWs, there appears an SIT frequency gap in which $K$ is imaginary or even zero, indicating pulse stopping or total absorption. Those frequencies for which $K=0$ we call stopping points. For all pulses with overcritical widths, there are two such points. 

The appearance of the forbidden band in SIT soliton dispersion law substantially affects two main characteristics of SIT phenomenon: lossless transparency and pulse velocity slowing down - pulse delay. For pulses whose carrier wave frequencies lie inside the SIT gap, where $K$ is imaginary, propagation is forbidden. 

Also, within the transparency windows, pulse velocity as a function of pulse width substantially deviates from the typical behavior: slow delay observed in atomic gases \cite{sit1,sit2,sit3,sit4,sit5}. To illustrate that, we have plotted a set of curves $V(\tau_0; X)$ for a few values of normalized frequency ($X=\omega/\Delta$) from both TWs -- Fig.\ref{fig2}. Thus, picking the value of $\omega/\Delta$ within the low-frequency TW but, bellow the one corresponding to the maximum of $K(X;\tau_0)$, curve $V$ versus $\tau_0$ exhibits behavior quite different from that encountered in studies of SIT in vapors and atom gases. In particular, as depicted by the dark green curve with circles in Fig. (\ref{fig2}), instead of gradually slowing down as a function of $\tau_0$, pulse velocity very steeply grows from zero and suddenly saturates at some constant value. 
For those $\omega/\Delta$ exceeding the one corresponding to the maximum of $K(X,\tau_0)$ pulse velocity smoothly grows from $V(\tau_0=0)$ up to the maximal value, after which it gradually decays and attains saturation value -- gray curve with circles. 

In the high-frequency TW, pulse velocity gradually decays as a function of the pulse width. Such behavior resembles that in the case of SIT in atomic gases. Nevertheless, due to the appearance of the photonic gap, here we predict some substantial differences. The first one is the appearance of an upper limit on pulse velocity $v_0\sim c_0\sqrt{1-\omega^2_0/\Delta^2}$. Next, pulse stopping appears for the frequencies corresponding to both stopping points. For example, for a pulse of the width $\tau_0=5$, these points are $\omega/\Delta\approx 1.07$ and $\omega/\Delta=1.223$. Pulse velocity delays for these two frequencies are depicted in Fig.2 by lines: blue one with circles ( $V(\tau_0; X= 1.07)$) and $V(\tau_0; X= 1.223)$ blue line. Finally, for all frequencies exceeding the one of higher stopping point, velocity versus pulse width tends to a saturation value corresponding to a limit $\tau_0\rightarrow \infty$ in equation (\ref{v}) while the asymptotic decay towards zero appears only in the resonant case $X=1$.

\begin{figure}[h]
\includegraphics[height=6.5cm]{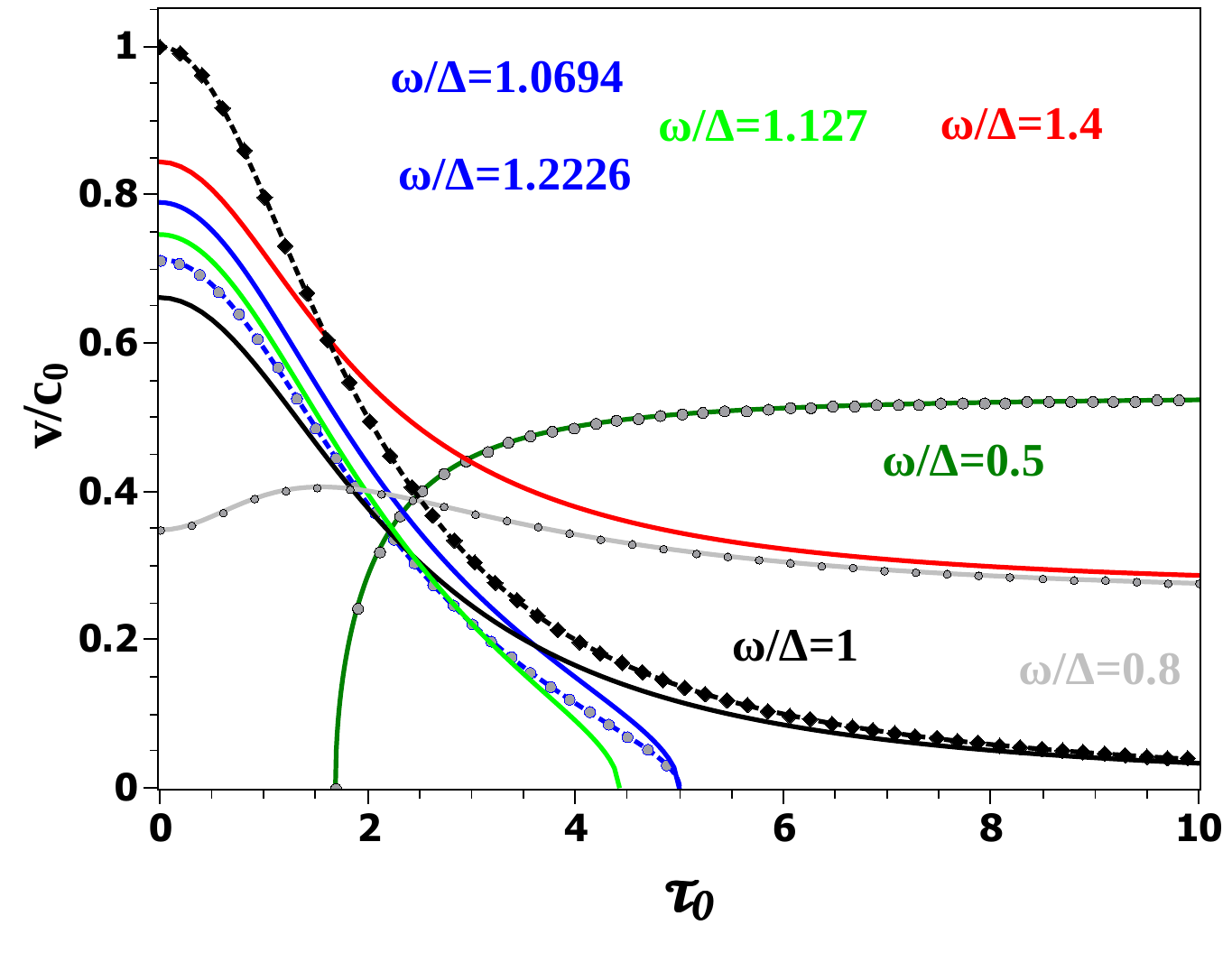}
\caption{SIT pulse velocity vs normalized pulse width -- $\tau_0$ for normalized gap $\alpha = 0.75$. Atypical behavior (the increase of $V(\tau_0; X)$ vs $\tau_0$) is observed for frequency ratio ($X$) from the low--frequency TW-- full lines with circles, olive and gray. Typical behavior (pulse velocity delay) appears for the frequency ratio from the high-frequency TW. Pulse stopping corresponds to points where $K(X,\tau_0)=0$. The full black line depicts the velocity delay of the resonant pulse - ($X=1$). For the comparison with SIT in atomic gases, we have plotted velocity delay for resonant pulse $X=1$ in the absence of a photonic gap. Black dotted curve with diamonds. }\label{fig2} \label{fig2}
\end{figure}
\section{concluding remarks}
We have demonstrated a possible emergence of the tunable SIT in a QMM comprised of an infinite 1d coupled cavity array filled by identical qubits. The proposed setup can be realized by embedding a row of periodically arranged local defects and vacancies in 2d photonic crystal  \cite{yariv,ccapolarit,sol4,sol5}. Their periodic variation may be controlled even tailored by varying photonic band gap frequency $\Omega=\omega_0-2J/\hbar$ providing, at the same time, the control over an effective qubit -- photon coupling constant $h\sim g\Omega$. SIT--soliton properties, its propagation in particular, are determined by carrier wave dispersion law $k(\omega)$, which substantially depends on pulse width. In particular, the medium is opaque for very short pulses with carrier wave frequency below the photonic gap. When the pulse width exceeds the critical one, a twin TW separated by a finite band gap appears in the SIT pulse dispersion law. Thus, managing pulse dispersion through the pulse input parameters, carrier wave frequency $\omega$, and width $\tau_0$, soliton can be significantly slowed and eventually stopped, which is of interest for the development of quantum memory devices.

The confirmation of the present findings would be an experimental observation of characteristic predictions such as the appearance of the upper limit of SIT pulse, finite saturation velocity for frequencies within the transparency windows, and finally, pulse stopping for frequencies corresponding to edge points of transparency windows.

Finally, we note that the results of our analysis, obtained within the idealized setting of identical units, may be extended toward more realistic systems. For example, the effects of IH broadening may be accounted for straightforwardly starting from the present un - broadened ("sharp line") solutions \cite{sit2}.

\begin{acknowledgments}
We thank D. Kapor for the fruitful discussion and useful comments on the manuscript. 
This work was supported by the Ministry of Education, Science, and
technological development of the Republic of Serbia. Z.I. acknowledges support by the "Vin\"ca" Institute -- special grant No. 104-64-2/2022-020, dated 30.12.2022. We also acknowledge the co-financing of this research by the European Union and Greek national funds through the Operational Program Crete 2020-2024, under the call "Partnerships of Companies with Institutions for Research and Transfer of Knowledge in the Thematic Priorities of RIS3Crete", with project title "Analyzing urban dynamics through monitoring the city magnetic environment" (project KPHP1 - 0029067).
\end{acknowledgments}

\section{appendix: derivation of reduced MB equations}\label{app}

\subsection{Basic system: discrete equations and continuum approximation}\label{MB}
Combining equations (\ref{se}) and their complex conjugate, we have derived evolution equations for the "field"
$u_n(t)=\frac{1}{\sqrt{N}}\sum_q e^{iqnb}(\alpha_q+\alpha^*_{-q})$ and atom variables, components of Bloch vector, defined as expectation values of Pauli matrices in a vector of state (\ref{vs}): $S^{x,y,z}_n=\langle \Psi(t)|\sigma^{x,y,z}_n|\Psi(t)\rangle$.
First, we use the third equation of the system and its complex conjugate. By adding and subtracting them, we derived
\begin{eqnarray}\label{maxwft}
\non \ddot u_q+\omega^2_q u_q=-\sum_n\omega_q g^{-q}_nS^x_n,\\
\omega_q =\omega_0-\frac{2J}{\hbar}\cos qb.
\end{eqnarray}
Similarly, we find evolution equations for Bloch vector components ($ S^{x}_n=\varPsi^e_n\varPsi^{g*}_{n}+c.c, \;\;
S^{y}_n=i\left(\varPsi^e_g\varPsi^{g*}_{n}-c.c\right), \;\;
S^{z}_n=|\varPsi^e_n|^2-|\varPsi^g_{n}|^2$) employing the
first two equations in (\ref{se}):
\begin{eqnarray}\label{blochft}
\non \dot S^x_n=-\Delta S^y_n,\\
\non\dot{S^y_n}=\Delta S^x_n-\sum_qg^{q}_n u_q S^z_n\\
\dot{S^z_n}=\sum_qg^{q}_n u_qS^y_n
\end{eqnarray}
Now, transforming back to direct space, the last equation attains the following form
\begin{widetext}
\begin{eqnarray}\label{dismaxw}
\ddot u_n+\omega^2_0 u_n-\frac{2\omega_0 J}{\hbar} (u_{n+1}+u_{n-1})+ \left(\frac{2 J}{\hbar}\right)^2 \frac{(u_{n+2}+u_{n-2})}{2}=
-\left[\omega_0 g S^x_n-\frac{Jg}{\hbar}(S^x_{n+1}+S^x_{n-1})\right].
\end{eqnarray}
\end{widetext}
Finally,
taking the continuum limit following known ansatz
$\mathcal{A}_n(t)\rightarrow \mathcal{A}(x,t)$ and $\mathcal{A}_{n\pm 1}\approx \mathcal{A}(x,t)+\pm b\mathcal{A}'(x,t)+(b^2)/2\mathcal{A}''(x,t)\pm ...$, primes denote derivation over $x$, we have
\begin{eqnarray}\label{maxw}
\non && \ddot u(x,t)+\Omega^2 u(x,t)-c^2_0 u''(x,t)=\\ &&- g\Omega S^x(x,t)+\frac{gJb^2}{\hbar}(S^x)''\\
\non && \Omega=\omega_0-\frac{2J}{\hbar},\;\; c^2_0=\frac{2J\Omega b^2}{\hbar},
\end{eqnarray}
and
\begin{eqnarray}\label{blochcont}
\non && \dot S^x(x,t)=-\Delta S^y(x,t),\\
&&\dot S^y(x,t)=\Delta S^x(x,t)-g u(x,t)S^z(x,t)\\
\non &&\dot S^z(x,t)=g u(x,t)S^y(x,t).
\end{eqnarray}
From  (\ref{maxw}) immediately follows the propagation condition for photons: $\Omega >0$ or $\omega_0> 2J/\hbar$, otherwise $c^2_0$ would be negative. In further analysis, in the spirit of continuum approximation, we will neglect the last term in (\ref{maxw}) being of the order of the product $b^2$ and the second derivative of slow variable $S^x$.
\subsection{reduced MBE}\label{RMBE}
Now we proceed by employing slowly varying envelope and phase approximation (SVEA) and introduce the new, slow, dynamical variables
\begin{eqnarray}\label{svea}
\non u(x,t)&=& f(x,t)\cos \varPsi(x,t),\; \;\varPsi=kx-\omega t+\phi(x,t),\\
S^x(x,t)&=& \mathcal{S}^x(x,t)\cos\varPsi (x,t)+\mathcal{S}^y(x,t)\sin\varPsi(x,t),\;\; \; \\
\non S^y (x,t)&=& \mathcal{S}^y(x,t)\cos\varPsi (x,t)-\mathcal{S}^x(x,t)\sin\varPsi(x,t).
\end{eqnarray}
Here $f(x,t)$, $\phi(x,t)$, are slow envelope, phase, while $k$ and $\omega$ are wave--vector and frequency of the carrier wave. Similarly, we substitute the original components of the Bloch vector with the slow ones.

The first step is the substitution of slow variables (\ref{svea}) in equations of motion (\ref{maxw}) and (\ref{blochcont}). Then, due to the assumed slowness of new variables, we neglect all second-order spatial and time derivatives and products of small variables. In such a way (\ref{maxw}) became:
\begin{widetext}
\begin{eqnarray}
\left( \dot f(x,t)+\frac{kc^2_0}{\omega}f'(x,t) \right)\sin\varPsi + \left(\dot\phi(x,t)+\frac{kc^2_0}{\omega}\phi'(x,t)+G(k)\right)f(x,t) \cos\varPsi=- \frac{g\Omega}{2\omega}\left(\mathcal{S}^y \sin\varPsi+\mathcal{S}^x\cos\varPsi\right).
\end{eqnarray}
\end{widetext}
Here $G(K)=\frac{\Omega^2+c^2_0k^2-\omega^2}{2\omega}$.
Thus, equating terms next to $\sin\varPsi$ and $\cos\varPsi$ we obtain system (\ref{rmbe1}).
Similarly, substitution of (\ref{svea}) in (\ref{blochcont}) yields
\begin{widetext}
\begin{eqnarray}
&&\non(\dot{\mathcal{S}^x}+\dot\varPsi\mathcal{S}^y)\cos\varPsi+(\dot{\mathcal{S}^y}-\dot\varPsi\mathcal{S}^x)\sin\varPsi=-\Delta(\mathcal{S}^y\cos\varPsi-\mathcal{S}^x\cos\varPsi),\\
&&\non(\dot{\mathcal{S}^y}-\dot\varPsi\mathcal{S}^x)\cos\varPsi-(\dot{\mathcal{S}^x}+\dot\varPsi\mathcal{S}^y)\sin\varPsi=\Delta(\mathcal{S}^x\cos\varPsi+\mathcal{S}^y\cos\varPsi)-{g} f(x,t)\cos\varPsi\mathcal{S}^z\\
&&\dot{\mathcal{S}^z}={g} f(x,t)\cos\varPsi\left(\mathcal{S}^y\cos\varPsi
-\mathcal{S}^x\sin\varPsi\right).
\end{eqnarray}
\end{widetext}
Using the above system, we have derived equations for new (slow) Bloch vector components in three steps. First, we multiply the first equation by $\cos\varPsi$ and the second one by $\sin\varPsi$. Then we subtract it from the first one. In the second step, we multiply the first equation with $\sin\varPsi$, the second one with $\cos\varPsi$, then we add them. In the final step, we perform averaging over the overall phase as follows
$\langle A(\Psi(x,t))\rangle_{\Psi} =\frac{1}{2\pi}\int^{2\pi}_0 d\Psi A(\Psi(x,t))$
These yields systems of equations of RMBEs:
pulse envelope ($f(x,t)$)) and phase ($\phi(x,t)$)
\begin{eqnarray}\label{rmbe1}
&&\dot f(x,t)+\frac{kc^2_0}{\omega}f'(x,t)=-\frac{g\Omega}{2\omega}\mathcal{S}^y(x,t),\\
\non &&\left(\dot\phi(x,t)+\frac{kc^2_0}{\omega}\phi'(x,t)+G(k)\right)f(x,t)=-
\frac{g\Omega}{2\omega}\mathcal{S}^x(x,t).
\end{eqnarray}
and the components of the Bloch vector:
\begin{eqnarray}\label{rmbe2}
\non && \dot{\mathcal{S}}^x=-(D+\dot\phi)\mathcal{S}^y\\
&&\dot{\mathcal{S}}^y=(D+\dot\phi)\mathcal{S}^x-\frac{g}{2}f(x,t)\mathcal{S}^z(x,t)\;\;\\
\non &&\dot{\mathcal{S}}^z=\frac{g}{2}f(x,t)\mathcal{S}^y(x,t), \;\; D=\Delta -\omega.
\end{eqnarray}

\subsection{Traveling wave solutions }\label{tvsol}
Introducing a time $\tau=t-\frac{x}{v}$ in a frame moving with velocity $v$ these two systems simplify to
\begin{eqnarray}\label{rmbe11}
\non &&f_{\tau}(\tau)=-\frac{g\Omega}{2\omega\gamma}\mathcal{S}^y(\tau),\\
&&\left(\phi_{\tau}(\tau)+\frac{G(k)}{\gamma}\right)f(\tau)=-\frac{g\Omega}{2\omega\gamma}\mathcal{S}^x, \;\;\\
\non &&\gamma=1-\frac{v_p}{v}, \;\; v_p=\frac{c^2_0}{\omega}k
\end{eqnarray}

\begin{eqnarray}\label{rmbe22}
\non && {\mathcal{S}}^x_{\tau}(\tau)=-(D+\phi_{\tau})\mathcal{S}^y\\
&&{\mathcal{S}}^y_{\tau}(\tau)=(D+\phi_{\tau})\mathcal{S}^x-\frac{g}{2}f(\tau)\mathcal{S}^z(\tau)\;\;\\
\non &&{\mathcal{S}}^z_{\tau}(\tau)=\frac{g}{2}f(\tau)\mathcal{S}^y(\tau).
\end{eqnarray}
Using the first equation from (\ref{rmbe11}) and the third one from (\ref{rmbe22})
we obtain differential equation for $\mathcal{S}(\tau)$:
\begin{equation}\label{deq}
\frac{d\mathcal{S}^z(\tau)}{d \tau}
=-\frac{g\gamma}{g\Omega}f(\tau)\frac{d f(\tau)}{d \tau}.
\end{equation}
Its integration with boundary conditions $\mathcal{S}^z(-\infty)=S_0$ and $f(-\infty)=0$, yields
\begin{equation}\label{sz}
\mathcal{S}^z(\tau)=S_0-\frac{\omega\gamma}{2\Omega}f^2(\tau).
\end{equation}
We now derive another conservation law important for further analysis, the evaluation of phase ($\phi$), in particular. For that purpose, we first derive an auxiliary relation using the first two equations from (\ref{rmbe11}) and (\ref{rmbe22}):
\begin{eqnarray}\label{sx}
\mathcal{S}^x_{\tau}=\frac{2\omega\gamma}{g\Omega}(D+\phi_{\tau})f_{\tau}(\tau).
\end{eqnarray}
Next, differentiating with respect to $\tau$ the second equation in (\ref{rmbe11}) and using the last result to eliminate $\mathcal{S}^{x}_{\tau}$ from it, we obtain: $\phi_{\tau,\tau}+2f_{\tau}\phi_{\tau}+(G/\gamma+D)f_{\tau}=0$.  After multiplication with $f$ last equation may be easily recast in the following form:
\begin{eqnarray}\label{phi}
\frac{d}{d\tau}\left\{f^2(\tau)\left[\phi_{\tau}(\tau)+\frac{1}{2}\left(\frac{G(k)}{\gamma}+D\right)\right]\right\}=0,\;\;\;
\end{eqnarray}
implying that:
\begin{eqnarray}\label{phi1}
f^2(\tau)\left[\phi_{\tau}(\tau)+\frac{1}{2}\left(\frac{G(k)}{\gamma}+D\right)\right]=const.\end{eqnarray}
Due to soliton boundary conditions $f(\pm \infty=0)$, the integration constant in the last equation is zero, implying that $\phi_{\tau}=const$ and, without loss of generality, may be set to be equal to zero. This yields the following result:
\begin{equation}\label{disp}
\frac{G(k)}{\gamma}+D=0,
\end{equation}
which we have used to evaluate a carrier wave quasi--momentum (\ref{k}).

Now we derive the differential equation for pulse amplitude. In the first step we substitute $\phi_{\tau}=0$ in (\ref{rmbe11}) and (\ref{rmbe22}). Next, we take the derivative over ($\tau$) of the first equation in (\ref{rmbe22}), then employing (\ref{sz}) and (\ref{sx}) to eliminate Bloch vector components $\mathcal{S}_x$ and $\mathcal{S}_z$, we obtain second order nonlinear differential equation for pulse amplitude:
\begin{eqnarray}\label{nlde}
f_{\tau\tau}+\left(D^2-\frac{g^2\Omega}{2\omega\gamma}S_0\right) f+\left(\frac{g}{2}\right)^2 f^3=0.
\end{eqnarray}
Imposing soliton boundary conditions, its first integral reads:
\begin{equation}
f^2_{\tau}(\tau)=\frac{g}{4}f^2(\tau)[f^2_0-f^2(\tau)],\;\; f^2_0=\left(\frac{4}{g}\right)^2\left(\frac{g^2\Omega S_0}{4\omega\gamma}-D^2\right).
\end{equation}
accordingly, the pulse envelope has a soliton-like form
\begin{eqnarray}\label{env}
f(\tau)=\frac{f_0}{\cosh\frac{\tau}{\tau_p}}, \;\; \tau_p=\frac{4}{f_0g}.
\end{eqnarray}
Substitution of the last result in (\ref{sz}) and (\ref{sx}) we found components of the Bloch vector.
\begin{eqnarray}\label{sol1}
\non \mathcal{S}^x(\tau)=\frac{2S_0D\tau_p }{1+(D\tau_p)^2}\mathrm{sech}(\frac{\tau}{\tau_p}),\\
\mathcal{S}^y=-\frac{2S_0}{1+(D\tau_p)^2}\mathrm{sech}(\frac{\tau}{\tau_p})\tanh(\frac{\tau}{\tau_p}),\\
\non \mathcal{S}^z=S_0\left(1-\frac{2}{1+(D\tau_p)^2}\mathrm{sech}^2(\frac{\tau}{\tau_p})\right)
\end{eqnarray}
Here for the practical calculations, we have used an alternate form for pulse amplitude:
\begin{eqnarray}\label{ampl}
f^2_0=\frac{4\Omega S_0}{\omega\gamma}\frac{1}{1+(D\tau_p)^2}
\end{eqnarray}
\subsection{Trigonometric parametrization and sine--Gordon equation}\label{tga}
Equivalence between reduced MBEs and the sine-Gordon equation holds in the resonance ($\Delta=\omega$) and for constant phase ($\dot\phi=0$). It yields $\mathcal{S}^x=const\equiv 0$ so that system simplifies to two first-order differential equations. They may be solved using trigonometric parametrization:
$\mathcal{S}^z=S_0\cos\theta(x,t), \;\; \mathcal{S}^y=S_0\sin\theta(x,t)$, where $\theta(x,t)$ denotes a Bloch angle. Their substitution into simplified system yields:
\begin{eqnarray}\label{angle11}
\dot\theta(x,t)&=&-\frac{g}{2}f(x,t),
\end{eqnarray}
while, in accordance with definition\cite{sit1,sit4}, pulse area is
\begin{eqnarray}\label{angle22}
\Theta(x)=\int_{-\infty}^{\infty}\dot\theta(x,t) dt &\equiv& -\frac{g}{2}\int_{-\infty}^{\infty}f(x,t) dt.
\end{eqnarray}

For finite detunings ($D$) factorization \textit{ansatz} \cite{sit1,sit4}: $\mathcal{S}^y(D)=F(D)\mathcal{S}^y(D=0) \Leftrightarrow\mathcal{S}^y(D)=F(D)\sin\theta)$, $\mathcal{S}^z=\mathcal{S}^0(1-F(D)+F(D)\cos\theta)$, is proposed. Here $F(D)$ is a spectral function whose explicit form we found by direct comparison of $\mathcal{S}^y(D)$, the second relation in (\ref{sol1}), with $\mathcal{S}^y(D=0)$. This yields: $F(D)=1/(1+(D\tau_p)^2)$.
Substitution
such trigonometric factorization \textit{ansatz} in (\ref{rmbe1}) and (\ref{rmbe2}) and employing the first relation from (\ref{angle11}) we finally derive the sine--Gordon equation for Bloch angle
\begin{eqnarray}\label{sge}
\left(\frac{\partial^2}{\partial t^2}+\frac{kc^2_0}{\omega}\frac{\partial^2 }{\partial x\partial t}\right)\theta(x,t)=\frac{g^2\Omega F(D)}{4\omega}\sin\theta(x,t).
\end{eqnarray}
Going over to a moving frame enables its integration resulting with:
\begin{equation}\label{thet}
\theta(\tau)=4\arctan(-\frac{\tau}{\tau_p}).
\end{equation}
Here $\tau_p= \sqrt{4\omega\gamma/(g^2\Omega F(D))}$ is pulse duration time or width in the time domain. Solution for pulse envelope, consistent with the one found above (\ref{env}), follows by differentiating the last expression over $\tau$.

\subsubsection{Derivation of energy balance equation}\label{enbal}
To prove lossless propagation, we use the energy balance equation, which we derive
employing the first equation in (\ref{rmbe1}) where we multiply both sides with $f(x,t)$. In the next step we eliminate $f(x,t)\mathcal{S}^y$ using the third equation from (\ref{rmbe2}) and we arrive at
\begin{equation}
\left(\frac{\partial }{\partial t}+\frac{k c^2_0}{\omega}\frac{\partial}{\partial x}\right)f^2(x,t)=-\frac{g\Omega }{\omega}\frac{\partial \mathcal{S}^z}{\partial t}.
\end{equation}
In the final step, we perform integration over time and obtain
\begin{eqnarray}\label{lossless1}
\frac{\partial W(x)}{\partial x}=-\mathcal{S}^0\frac{F(D)g\Omega }{kc^2_0}\bigg(1-\cos\theta(x)\bigg).
\end{eqnarray}
Here $\mathcal{S}_0=\langle\Psi(-\infty)|S^z_n|\Psi(-\infty)\rangle\equiv |\varPsi^e_{n}|^2-|\varPsi^g_{n}|^2$ represents the initial value for population inversion. It may take values $\mp 1$ depending on whether all "atoms" are in the ground (absorbing media) or excited (amplifying media) state.

Pulse energy density is $W(x)\sim \int_{-\infty}^{\infty} dt u^2(x,t)$. Following SVEA, pulse energy density averaged over the overall phase attains the form $W(x)\sim \frac{1}{2}\int_{-\infty}^{\infty}dt f^2(x,t)$.
%%%%%%%%%%%%%%%%%%%%%%%%%%%%%%%%%%%%%%%%%%%%%%%%%%%%%%%%%%%%%%%%%%%%%%%%%%%%%%%%

%%%%%%%%%%%%%%%%%%%%%%%%%

%%%%%%%%%%%%%%%%%%%%%%%%%%%%%%%%%%%%%%%%%%%%%%%%%%%%%%%%%%%%%%%%%%%%%%%%%%%%%%%%
%%%%%%%%%%%%%%%%%%%%%%%%%%%%%%%%%%%%%%%%%%%%%%%%%%%%%%%%%%%%%%%%%%%%%%%%%%%%%%%%
\end{document}